%% file: ms.tex
\documentclass[10pt]{article}
\pdfoutput=1

\usepackage[papersize={8.5 in, 11 in}, nohead, includeheadfoot, left=1.5 in, right=1.5 in, vmargin=1 in]{geometry}

\input{packages.tex}

\input{math.tex}

\input{misc.tex}

\usepackage[backend=bibtex, defernumbers = true, style = authoryear, uniquename = false]{biblatex}
\title{Regression Analysis of Proportion Outcomes with Random Effects}

\author{Colman Humphrey \quad Daniel Swingley \\\textsc{\small University of Pennsylvania}}
\date{}

\begin{document}

\maketitle

\begin{abstract}
A regression method for proportional, or fractional, data with mixed effects is outlined, designed for analysis of datasets in which the outcomes have substantial weight at the bounds.  In such cases a normal approximation is particularly unsuitable as it can result in incorrect inference.  To resolve this problem, we employ a logistic regression model and then apply a bootstrap method to correct conservative confidence intervals.  This paper outlines the theory of the method, and demonstrates its utility using simulated data.  Working code for the R platform is provided through the package glmmboot, available on CRAN.
\end{abstract}

\section{Introduction}

Some processes yield bounded outcomes with heavy representation at the
bounds. For example, if we consider the number of pages of a given
issue of a magazine that are read by its subscribers, the distribution
likely includes a large number of zeros (from those who didn't have
time to open the issue at all) and a large number at the maximum
(from those who read cover to cover), and then some proportion spread
in between. Or, to take a second example, a sampled person's
proportion of nights spent in a hospital in a given week must fall
between zero and one, but among persons there will usually be vastly
more zeros than anything else (most people aren't hospitalized) and
otherwise a disproportionate number of ones (some hospitalized people
spend the whole week there, in longer-term care). Our paper discusses
a technique for the analysis of such datasets in which the boundary
values are common but don't exhaust the set. Formally, we are
interested in outcome data $Y \in [0,1]$, with many data points at the
endpoints $\{0,1\}$. With trial-level covariates, we wish to model
these data with regression, rather than collapsing to group means. Our
covariates include both fixed effects and random effects.

If our outcomes were exclusively in $\{0,1\}$, we could for example run
a straight-forward logistic regression with mixed effects. If our
outcomes were strictly in $(0,1)$, we could consider a beta regression or
similar methods. But these do not fit the problem at hand.

In the data we consider here, we do not feel that the zeroes and ones
are conceptually different to the data within the interval. We would
consider $0$ and $0.01$ almost the same. Similarly, seeing repeated
measurements from a given input\footnote{That is, the covariates are the
 same, $\x_1 = \x_2$} of $(y_1 = 0.8,\; y_2 = 1)$
would not be far from $(y_1 = 0.88,\; y_2 = 0.92)$.
Thus we are averse to methods that fit separate covariates or models for the
endpoint cases and the continuous cases, i.e. $\{0,1\}$ vs.\ $(0,1)$.

\section{Motivating Problem}

While our method is more general than will be outlined in this
section, we use this simple case to illustrate it.

\subsection{Data Description}

We consider outcome data of the form $Y_i \in [0,1]$, for $i = 1,\hdots,n $, with a covariate
of primary interest, $X_i \in \{0,1\}$, which can be thought of as
a treatment indicator, along with fixed-effect covariates not of primary
interest. Our data also contains a random effect $Z_i \in \{1, 2,\hdots K \}$. We
could think of this for example as $K$ participants each being measured over
several trials.

\subsection{Conditional Means}

With the shorthand $Y[\bm{x} = 1]$ meaning the $Y$ values for which $X
= 1$, a typical conditional mean comparison would compare
$Y[\bm{x} = 1]$ to $Y[\bm{x} = 0]$, i.e. comparing the means of each
condition, with suitable standard deviations. Or we could do a
comparison within the levels of our random effect, by comparing:
\begin{align}
  \begin{split}
\left\{
\begin{array}{ccc}
  Y[\bm{x} = 1, \bm{z} = 1]& \text{vs} &Y[\bm{x} = 0, \bm{z} = 1] \\
  Y[\bm{x} = 1, \bm{z} = 2]& \text{vs} &Y[\bm{x} = 0, \bm{z} = 2] \\
  &\vdots& \\
  Y[\bm{x} = 1, \bm{z} = K]& \text{vs}  &Y[\bm{x} = 0, \bm{z} = K]
\end{array}
\right.
  \end{split}
  \end{align}

And of course we can have multiple covariates of interest, which
are not restricted to binary variables.

One problem with collapsing over trials as in the above is that we lose the ability to take
trial-level predictors into account.  Once we average over trials, we lose this small grain size in the predictors. This will affect our estimate of the effect of $\bm{x}$, and
will not allow us to understand what role the other covariates play.

In the case of purely binary data, \textcite{dixon2008models} points
out three further disadvantages of modelling constrained data in this
way: if we wish to separate by some factor, we may have few
observations in each condition, and the normal approximation could be
inaccurate; due to the constraint, we may have floor or ceiling
effects from averaging; and averaging artifacts can arise due to different
numbers of observations across subjects or conditions. These arguments
also apply to proportion data. Thus, we are seeking a method that allows us to retain the individual trial as the basic unit of the analysis, in spite of the fact that the trial-level outcomes are neither normal nor binary and therefore violate the assumptions of many familiar techniques.

\section{Regression}

The goal of regression is to model the expected value of $Y_i$ as a
function of our covariates $\x_i = (x_{i0},\; x_{i1},\;\hdots\; x_{ip})'$ to get
$\E[Y_i \mid \x_i]$. For now, for the sake of simplicity we will ignore random
effects.

\subsection{Linear Modelling}

In a linear model, we assume the expected value of $Y_i$ is a linear combination
of the covariates, i.e. we have a vector $\bm{\beta}$ such that:
\begin{align}
  \E[Y_i \mid \x_i, \bm{\beta}] = x_{i0} \beta_0 + x_{i1} \beta_1 + \hdots + x_{ip} \beta_p =
  \x_i' \bm{\beta}
\end{align}

We can also think of this expectation not just as
a linear combination of the covariates, $\x_i' \bm{\beta}$, but as a function of this combination:
the identity function $I(x) = x$. That is, we can write:
\begin{align}
  \label{eq:exp}
  \E[Y_i \mid \x_i, \bm{\beta}] = I(\x_i' \bm{\beta}) = \x_i' \bm{\beta}
\end{align}
This doesn't change anything in this context, but using a function other than the
identity would change the model.

Using a linear model with proportional outcomes raises many issues,
similar to the problems that would be caused by
using linear modelling with binomial $Y$. Most directly, due to the bounding of $Y$, we do not
expect each covariate to affect the expected value of $Y$ the same way
for different baseline levels of $\E [Y]$; for example, when $Y$ is expected to be large versus when it is
  expected to be close to $\sfrac{1}{2}$. Similarly for an unbounded
covariate $x_j$, we don't expect the effect of $x_j$ on $Y$ to remain
the same for small and large values of $x_j$. There is no
guarantee that predictions will be contained in the interval $[0,1]$,
and the usual methods for fitting linear models assume the wrong error
structure, especially for $Y$ values near the boundaries.

\subsection{Logistic Modelling}

One way to fix these problems is
to change the function of the linear combination of variables $\x_i' \bm{\beta}$.
Generalized linear models, or GLMs are one way to do this.  GLMs are a flexible set of models that allow a large
variety of outcome distributions.

Similar to \eqref{eq:exp}, we'll have:
\begin{align}
  \label{eq:Gexp}
  \E[Y_i \mid \x_i, \bm{\beta}] = G(\x_i' \bm{\beta})
\end{align}
with some function $G(.)$ instead of the identity $I(.)$ from before.
We need $G(x)$ to map to $(0,1)$,\footnote{Zero and 1 are excluded because 
 $Y$ would never have an expected value equal to either of the boundary values.} and
it should be increasing: as $\x_i' \bm{\beta}$
increases, so does $\E[Y_i \mid \x_i, \bm{\beta}]$. $G$ links
the predictors to the mean.  It is what is usually referred to as the \textit{inverse link function} in the terminology of generalized linear modeling.
  
A common and often recommended choice of linking function is
\[
G(x) = \frac{e^x}{1 + e^x}
\]

This gives logistic regression. With binary $Y$, this linking function has the
interpretation of modelling the log odds as a linear combination of
the covariates. While we don't maintain that interpretation, we can
still think of modelling proportion ratios instead of proportions:
\begin{align}
  \label{eq:logprop}
  \log \left(\frac{\E[Y_i \mid \x_i, \bm{\beta}]}{1 - \E[Y_i \mid \x_i, \bm{\beta}]}\right)
  = \x_i' \bm{\beta}
\end{align}

By employing the logistic function, we remedy the problem of ceiling
effects. For example, an expected value for $Y$ of $0.99$ is as
``far'' from $0.95$ as $0.95$ is from $0.78$.\footnote{This property of the logistic transform is effective in keeping predictions within bounds, but whether it is appropriate in considering the real-world domain of the analysis is important for the modeller to bear in mind. There may be domains in which a change from 50\% to 55\% is, in fact, the same as a change from 90\% to 95\% in how the predictors influence outcomes.  This is a question that could be addressed empirically, but is not one that we evaluate here.}
This method forms the basis of the method from
\textcite{papke1996econometric} to model $Y \in [0,1]$.

\subsection{Quasi-Likelihood Methods} \label{quasill}

In the binary case of logistic regression, $\E[Y] = \P(Y = 1) \cdot 1
+ \P(Y = 0) \cdot 0 = \P(Y = 1)$, directly modelling the probability
of $Y$. This makes solving for $\bm{\beta}$ simple, because we have
the full distribution of the data, and therefore the likelihood. We
generally solve for $\bm{\beta}$ by maximising the log likelihood
$l(\bm{b})$:
\begin{align}
  \begin{split}
  l(\bm{b}) 
  =& \sum_{i} \log \P(Y_i = y_i \mid \bm{\beta} = \bm{b}, \bm{x_{i}})\\
  =& \sum_{i} l_i(\bm{b})
  \end{split}
\end{align}

We can use \eqref{eq:Gexp} to write:
\begin{align}
  \begin{split}
  l_{i}(\bm{b})
    =&\; y_{i} \log[G(\x_i' \bm{b})] + (1 - y_{i})
    \log\left[1 - G(\x_i' \bm{b})\right]
    \label{eq:lprobinterp}
  \end{split}
\end{align}
Full details are given in appendix \ref{appen_quasill}. 

We note that above log likelihood function can be evaluated with
$y_{i} \in [0,1]$, not just for $\{0,1\}$. It functions as linear
interpolation between the $0$ and $1$ cases.  This is no longer a true
likelihood, but a true likelihood is not required to find a consistent
estimate of $\bm{\beta}$.
From \textcite{gourieroux1984pseudo}, we only need \eqref{eq:Gexp}.

We maximise the full quasi-log likelihood and thus we find the
parameter vector that best fits the expected proportion $Y$ given our
predictors. The result is analogous to the outcome of an ordinary logistic
regression. Even though the outcome being modeled does not fall only
into values of zero and one, we still find the best fit for the
expected value. 

\subsection{Interpreting Logistic Models}

Interpretation of logistic models is less transparent than
interpretation of linear models. Let us assume we have fit our model
and have estimated $\hat{\bm{\beta}}$ as our best-fitting parameter
vector; and assume further we also have confidence intervals and
$p$-values.

Because the logistic is still a monotonic fit in our covariates,
we can continue to interpret a highly significant variable as meaning
the associated covariate is correlated with $Y$. For example, if $\beta_2 \gg 0$,
i.e. significantly above zero (with appropriately small p-value), we
know that $x_2$ is positively correlated with $Y$. That is, as $x_2$
increases, our expected value for $Y$ will increase, and we are confident
this is a significant pattern. This is the most fundamental and important
interpretation.

The direct rate of change using the derivative is:
\begin{align}
  \frac{\partial}{\partial x_j} \E[Y \mid \x, \bm{\beta}]
  = \beta_j \frac{\exp(\x_i' \bm{\beta})}{(\exp(\x_i' \bm{\beta}) + 1)^2}
\end{align}

This can be used for any specific $\x$, but is not easy to
interpret as-is. One simplification is that near the midpoint,
i.e. when $\E[Y \mid \x, \bm{\beta}]$ is close to $\sfrac{1}{2}$, this
derivative will be approximately $\sfrac{\beta_j}{4}$. So if $\beta_3
= 0.2$, then a unit increase in $x_3$ will increase the expected
proportion by approximately $0.05$. This interpretation only works for
small changes in $x$ and/or small values of $\beta_3$.
When the predictions are close to zero or one,
where the nonlinearity of the logistic linking function becomes more important,
the effects of a change in a variable on the mean of $Y$ will be smaller.

As mentioned in the prior section, in the binary case of logistic
regression we can interpret our coefficients in terms of odds:
the ratio of the probability of $Y = 1$ over the probability of $Y = 0$.
Instead of probabilities, we interpret our coefficients in terms of proportion
ratios. For example, if we're measuring the proportion of time a
football team has possession of the ball, an expected proportion of
$0.75$ is equivalent to expecting that team to have the ball three
times as much as the opposition, $\sfrac{0.75}{0.25} = 3$.

Letting $\widehat{\text{prop}}_i$ be the expected proportion in trial $i$, and
exponentiating both sides of \eqref{eq:logprop}, we can write:
\begin{align}
  \widehat{\text{ratio}}_i = \frac{\widehat{\text{prop}}_i}{1 - \widehat{\text{prop}}_i} = 
  \exp(\x_i' \hat{\bm{\beta}})
\end{align}

To understand what happens when we change a value of $\x$, let
$\x_i^\text{old}$ be a vector of covariates.
Assume we increase $x_3^i$ by $1$, to get $\x_i^\text{new}$.
These covariates, new and old, give different expected ratios,
$\widehat{\text{ratio}}_i^\text{new}$ and $\widehat{\text{ratio}}_i^\text{old}$.
We can look at how the ratios will change. First note that
$(\x_i^\text{new})' \hat{\bm{\beta}} = (\x_i^\text{old})' \hat{\bm{\beta}}
+ \hat{\beta}_3$, since we increase $x_3$ and leave the rest.
\begin{align}
  \begin{split}
    \frac{\widehat{\text{ratio}}_i^\text{new}}{\widehat{\text{ratio}}_i^\text{old}}
    &= \frac{\exp((\x_i^\text{new})' \hat{\bm{\beta}})}
         {\exp((\x_i^\text{old})' \hat{\bm{\beta}})} \\
         &= \frac{\exp((\x_i^\text{old})' \hat{\bm{\beta}} + \hat{\beta}_3)}
         {\exp((\x_i^\text{old})' \hat{\bm{\beta}})} \\
         &= \frac{\exp((\x_i^\text{old})' \hat{\bm{\beta}})\exp(\hat{\beta}_3)}
         {\exp((\x_i^\text{old})' \hat{\bm{\beta}})} \\
         &= \exp(\hat{\beta}_3)
  \end{split}
\end{align}

Thus we have an interpretation of $\exp(\beta_j)$ as the multiplicative
change in the ratio of proportions. Going back to the sports analogy, imagine that ball possession is affected by how much rain fell before the match. Suppose our coefficient for rainfall (in mm) was $0.4$. An extra mm of
rain would increase the ratio by a factor of $\exp(0.4) \approx 1.5$.
So if our team had been
expecting $30\%$ possession with 1 mm of rain, a ratio of $\sfrac{3}{7}$, 
they expect $1.5$ times this with 2 mm of rain, or a ratio of $0.64$. This corresponds to
$39\%$ possession.

Appendix \ref{sec:append_interp_log} offers interpretations
when $\E[Y \mid \x, \bm{\beta}]$ is close to the endpoints.

\subsection{Random Effects}

The preceding examples assumed relatively simple models, but for many
practical applications mixed-effects modeling is more appropriate.
Once mixed effects are considered, $\E[Y \mid \x, \bm{\beta}] =
G(\x'\bm{\beta})$ is no longer true. A typical mixed effects model
adds a design vector $\bm{z}$ and a random vector $\cB$\footnote{Note that $\cB$
  and $\bm{\beta}$ are not the same}. The
design vector is fixed.  It can represent, for example, 
a vector containing a unique identifier for each participant in a study,
while the random vector represents the variation from the random effects.
When adding random effects,
we can make a
similar statement about the conditional distribution of $Y$ given $\cB
= \bm{b}$:
\begin{align}
  \label{eq:RE_G}
  \E[Y \mid \x, \bm{\beta}, \bm{z}, \cB = \bm{b}] = G(\x'\bm{\beta} + \bm{z}'\bm{b})
\end{align}
with $\cB \sim \cN(0, \Sigma)$, a normal distribution with unknown variance.
The expected
value of $Y$ without the unknown random vector is the average value over
the distribution of $\cB$:
\begin{align}
  \label{quasi_glmm}
  \E[Y \mid \x, \bm{\beta}] = \int_b G(\x'\bm{\beta} + \bm{z}'\bm{b}) f(\bm{b}) \td \bm{b}
\end{align}
where $f(.)$ is the pdf of $\cB$. See \textcite{bolker2009generalized} for details.

For a given realization of random effects,
i.e. a fixed but
unknown vector $\bm{b}$ such that $\cB = \bm{b}$,
the conditional interpretations already discussed are
valid. For example, say we want subject-level random effects
(i.e. random intercepts). For a specific subject, we can interpret the
effects of $x$ as in the prior subsection.

\subsection{Summary}

In summary, to run a regression method on outcomes in $[0,1]$ with
trial-level covariates, we run a quasi-likelihood
version of logistic regression: we keep the interpretation
of fitting the mean, but no longer assume we're specifying the
whole distribution - in particular we don't need to specify the variance.
On top of this, we add random effects, for appropriately
modelling many practical applications. In section \ref{sec:practical_stuff},
we will describe current statistical software for fitting such models.

In some cases, no more needs to be added. A small change in model specification
allows non-binary models to be fit by methods traditionally associated
with binary data, with consistent parameter estimation and valid standard
errors. However, these models are conservative. We discuss the issue and a
remedy in the following section.

\section{Variance Correction}

Nearly all estimation methods for the parameter vector will
also produce standard errors. The specific calculation
will depend on the software implementation, but generally
the standard errors will be generated with respect to the
nominal likelihood. For example, if we assume a logistic
link function for our quasi-likelihood fit,
we may be given standard errors generated from
that functional form.

We only wish to specify the conditional mean of
our data, not the variance, thus we don't want to put any assumptions
on the variance. Further, typical implementations
for proportion or fractional data can be conservative,
i.e.\ the standard errors will be too large. This is partially due
to the standard errors being estimated assuming the worst
case scenario, in which the variance of $Y$ given $\x$ is $\nu(1 - \nu)$,
where $\nu = \E [Y \mid \x, \bm{\beta}]$. This is the correct specification for
binary $Y$, while proportional $Y$ has less variance on average, and thus more accuracy.

\subsection{Correction Methods}

There are three main ways to correct the variance estimates. The first is by using sandwich estimators, also known
as heteroscedasticity-consistent standard errors \parencite{white1980heteroskedasticity}.
\textcite{papke1996econometric} use sandwich estimators.
We don't use these because we don't wish to rely on asymptotics - there is no guarantee
that such asymptotics will converge in a reasonable time, especially
if any skewness is present. In our simulations, sandwich methods seemed to be conservative.

Secondly, we can use over-dispersion methods (\cite{mccullagh1984generalized}
page 124). These are hard to specify with random effects, and rely on the
functional form of the mean\footnote{That is, the relationship between the
  inputs and the mean.}, so we don't use these.  

Finally, we can use bootstrap methods, which we will describe further
in the following sections. 

\subsection{Bootstrap Method}
\label{bootmethod}

Bootstrapping is a resampling technique that uses resampling with
replacement. In inference, we're trying to use information from our
sample, i.e.\ our original data, to infer details about the population.
We can approximate this relationship by treating our sample as a
population, and letting our resamples act as samples from it. In other
words, our resamples are to our sample as our sample is to our
population. 

Consider, for example, estimating the variance of a coefficient.  A given regression analysis over a sample from a population will yield a coefficient specifying the influence of each predictor. This coefficient may be the closest estimate for the analyzed sample, but it would be good to have an idea of how representative that coefficient is over the whole set of samples we could have randomly drawn from the same population.  We can simulate this process of having more than one sample by bootstrapping.
We resample our data with replacement multiple times, and
estimate the coefficient for each resampled data set. We then use the
variance of these resampled coefficients as our
variance estimate for the sample.

For the quasi-likelihood method outlined in section \ref{quasill},
we work with the assumption outlined in \eqref{eq:Gexp}, i.e.\ that we're
fitting the expected value of $Y$. We fit this quasi-likelihood method
assuming $G$ is the logistic function. As discussed,
this implies a variance structure on $\bm{\beta}$,
but we might not trust that this process will generate correctly specified $t$
distributions for inference.
In particular, because binary data generally presents greater variance than the
continuous data we are contemplating here, there is a risk that the $t$ distributions will yield overly conservative inference, and thus less power.

\subsection{Bootstrap Resampling with Random Effects}
\label{set_resamples}

Ordinary bootstrap resampling is simple: resample each row with
replacement, with equal weight.  When random effects are included in a
model, we can no longer resample in such a straightforward
manner. Instead, we have to simulate the data-generating process as
closely as possible. We do this by block resampling \parencite{mccullagh2000resampling}:
we resample from blocks with replacement, where the blocks are defined
by the random effects.

For example, with subjects as our random effect,
we resample subjects with replacement. Every time a subject is selected,
we select all their observations, without replacement. \textcite{mccullagh2000resampling}
shows that resampling a second time within the blocks is conservative and unnecessary.

With one random effect, the above is well-defined. With two or more, a valid resampling procedure
becomes more difficult to define. We discuss in appendix \ref{multi_RE} how
to implement methods from \textcite{owen2012bootstrapping} if desired.

\subsection{Bootstrapping One Random Effect, not Multiple}

In practice, we don't use 
multiway bootstrap methods: they are generally even
more conservative than using the implied binary variance structure, i.e.
the raw model output from running the proportional models in most
statistical software.

Thus, in our analysis pipeline we enter all appropriate random effects into the original model; then, when using bootstrapping to estimate the variance of the coefficients, the bootstraps only include one of the random effects.
\textcite{bakshy2013uncertainty} show that taking into account
only one type of dependence structure, or random effect, gives
valid inference under the typical null hypothesis, even
without building a model with random effects. In our simulation examples,
taking only one random effect into account on the level of the bootstrap retains
correct (or conservative) coverage.

As we discuss in our simulation study, we bootstrap over
the random effect that has more levels because the coverage
of our intervals is much better with more levels\footnote{In
  practice, we actually use the random effect with the largest entropy,
  $H(X) = - \sum p(x) \log p(x)$. Generally this coincides with
  the effect with most levels.}.

\subsection{Generating Bootstrap Results}
\label{sec:gen_boot_results}

Once we've chosen what level we're resampling with replacement from,
we resample $R$ times for some
large $R$ and refit our model on each new dataset. This generates a
set of parameter vectors, $\hat{\bm{\beta}}^*_1, \hat{\bm{\beta}}^*_2,
\hdots, \hat{\bm{\beta}}^*_R$, along with standard error vectors from
each bootstrap, which we'll call $\hat{\bm{se}}^*_{\beta_j}$, for $j =
1, \hdots, R$.  From our regression on the full original dataset,
which we'll call the base regression, we have $\hat{\bm{\beta}}$ and
$\hat{\bm{se}}_{\beta}$.

Note also that our boostrap methods are non-parametric: we resample
blocks of trials and refit our model on the resampled data, rather than resampling
residuals. The alternatives, bootstrapping residuals, are difficult to
justify since we don't specify a model, only a mean structure.
With our non-parametric bootstrap, we must
be wary of small-sample problems with discrete covarites.

It is non-trivial to generate confidence intervals and
p-values from boostrap estimates. We follow \textcite{hesterberg2015teachers}
in using our boostrap samples to approximate the distribution of the
``$t$'' values produced by our regression, from
\textcite{efron1994introduction}. This statistic is a pivotal
quantity: it does not depend on unknown parameters, such as the model variance\footnote{Not strictly
  true in random effect models, but the difference is small enough to not
  matter.}.

Suppose that we're interested in the $j$th covariate parameter, $\beta^j$. We
have the estimate $\hat{\beta}^j$, and the standard error
$\hat{se}^j_{\beta}$ from our base regression. (For now, we drop the
$j$ notation.) In testing against zero, we typically generate:
\begin{align}
  t_\text{plain} = \frac{\hat{\beta} - 0}{\hat{se}_{\beta}}
\end{align}

Under any issues outlined in section \ref{bootmethod}, we might
not believe that this statistic will actually have a $t$ distribution.
Instead of assuming a distribution on this statistic, we use our bootstrap
samples to estimate the distribution:
we generate the $R$ bootstrap $t$ statistics in the same way as
we do in the base regression:
\begin{align}
  t^* = \frac{\hat{\beta^*} - \hat{\beta}}{\hat{se}_{\beta^*}}
\end{align}
Letting $q_\alpha$ be the $\alpha$ quantile of the bootstrap $t$
distribution, the interval is then:
\begin{align}
  (\hat{\beta} - q_{1 - \sfrac{\alpha}{2}} \hat{se}_{\beta},\;
  \hat{\beta} - q_{\sfrac{\alpha}{2}} \hat{se}_{\beta})
\end{align}

Just as with any confidence intervals from regression,
we can reject e.g.\ zero at level $\alpha$. We can turn this
around to form p-values: we calculate the largest $\alpha$
such that this interval contains zero.

\subsection{Confidence Intervals and $p$-values for Random Effects}

Non-pivotal bootstrap methods can suffer from two sources of narrowness.
They come from the equivalent of using
$z$ instead of $t$, and from dividing
the sample standard deviation by $\sqrt{n}$ instead of $\sqrt{n - \text{df}}$ 
to get standard errors.
The bootstrap $t$ method
from the previous section in theory does not suffer from these bootstrap issues,
since it directly estimates the $t$-distribution, specifically the
$t$-distribution with the correct degrees of freedom.

Many statisticians who are well-versed in random effects modeling are
wary of constructing $p$-values for models with random effects; see
for example \textcite{baayen2008mixed}.
We are similarly cautious. Our bootstrap method is 
capable of providing $p$-values if desired, warnings
notwithstanding.

See the next section for advice about interpreting $p$-values less than
$0.05$. 

\subsection{Practical Considerations for Bootstrapping}

\textcite{hesterberg2015teachers} recommends $R > 15,000$ for published works, and we agree.

If time is a concern, and e.g. model-fitting is slow for a given
problem, doing $15,000$ runs multiple times may not be feasible.  For small-scale model checking,
the standard errors straight from the base
regression are likely to give an acceptable first approximation.
Of course these multiple models should not be guiding your decision of model selection.

In terms of the $p$-values: this large $R$ is recommended because it gives
reasonable guarantees about accuracy near the typical $0.025$ quantile points.
If we're estimating a smaller $p$-value than $0.05$, and are therefore outside
of the $0.025$ bound on either side, our accuracy will be lower, and
our $p$-value estimate will be subject to greater random noise due to
resampling for the bootstrap.  This is because we are
estimating the $p$ value by counting how often a rare event occurs.
The number of sample draws thus has to be very large to accurately
estimate the long tails of the probability distribution.
Intuitively, you cannot accurately judge the actual probability of a
one-in-a-hundred event by looking at a hundred samples, or a thousand;
as we walk down the tail of the distribution, the granularity of the bootstraps overwhelms their accuracy.
We therefore advocate against taking $p$-values below $0.05$ as being informative.
If the researcher is particularly interested in a threshold below $0.05$,
\textcite{hesterberg2015teachers} provides the calculation for how many
resamples gives acceptable accuracy.
In any case, using miniscule $p$ values as shorthand for strong effects is contraindicated
as a general matter, not just for bootstrapping analyses (see for example \textcite{sullivan2012using}), but this practice carries additional risks in bootstrap analyses.

\section{Work Flow}
\label{sec:practical_stuff}

Before we bootstrap, we need an algorithm capable of solving the quasi
general mixed model from \eqref{quasi_glmm}.  We use the
\texttt{glmmTMB} \textbf{R} package \parencite{R2017glmmTMB}. For those familiar with the \textbf{R}
package \texttt{lme4}, the notation is the same.

While this package does not contain an equivalent of the \emph{glm}
function's \emph{quasibinomial} distribution family, we can run it
with the standard \emph{binomial} family input (with default logistic
link). This gives us the solution under equation \eqref{eq:Gexp} as
desired, although with conservative standard errors referred to in
section \ref{bootmethod}.

Once the practitioner is satisfied that her model runs as desired in
\texttt{glmmTMB}, she can use the bootstrap to generate confidence
intervals and $p$-values for the parameters in her model. We call the
model run on the original data the ``base model''. Code for the
following work is provided on CRAN, at \texttt{https://cran.r-project.org/web/packages/glmmboot}.

The number of bootstrap resamples, $R$, is set to some reasonable
choice ($\sim 15,000$ per Hesterberg). For each sample, a resampling
of the random effects is generated as in section \ref{set_resamples},
which generates a full resampled data set. The model is then run
exactly as in the base model on this resampled data set. The output is
saved, i.e.\ estimates and standard errors from each bootstrap run.  This
is by far the longest step in terms of computation time.
Once these are run and saved, all bootstrap output plus the output
from the base model are combined to form 
produce corrected confidence intervals and p-values,
using the bootstrap $t$ method, from section \ref{sec:gen_boot_results}.

\subsection{Recommended Output}

Our next task is to produce overall output from the method, i.e.\
the result of the full work flow from above.

We don't recommend using the bootstrap to adjust the base model's
parameter estimates, $\hat{\bm{\beta}}$. This is the most important
model output, and its value and interpretation does not depend on
bootstrap output.   We do recommend reporting the bootstrap
confidence intervals. The bootstrap $p$-values are usually
of interest too.

An optional but interpretable output is the standard errors from the
base model. Since these are generally conservative and are independent
of the bootstrap, they provide a grounding for significance results
from the bootstrap, as well as a quick indication of how much less
conservative the bootstrap approach is.  Further, the bootstrap CIs
are not necessarily symmetric around the estimates, so these
unadjusted standard errors can be useful for a quick symmetric
estimate of variability.

\section{Simulation Study}

We run a factorial simulation study over multiple parameters including
the number of rows, 1000 and 2000; the number
of fixed effect parameters (not including the intercept), 3 and 10;
the number of random effects, one or two; crossed and uncrossed random
effects; correlated and uncorrelated parameters.
Further, we vary the number of levels of each random effect,
either 10, 20, 40 or 80. In half our simulations, we use zero
fixed effects, and random non-zero effects in the other half.

We generate a random vector $\bm{\cB}$ from a normal distribution.
The known design matrix $X$ is generated according to a mixture
of a randomly correlated multivariate normal distribution and
independent uniform random variables. When the fixed effects
are non-null, we generate normal random fixed effects. 
 
The mean vector is then fit according to equation \eqref{eq:RE_G},
and finally $Y$ is generated in one of two ways. Mostly, we
generated from a beta distribution, with the parameters set to:
\begin{align}
  \begin{split}
    \alpha &= \mu^2 (1 - \mu)\sigma^{-2} - \mu \\
    \beta &= \alpha \left(\mu^{-1} - 1\right)
  \end{split}
\end{align}
$\mu$ is from the mean vector. We set $\sigma^2$ to be
$\rho^2 \mu(1 - \mu)$, with $\rho \in (0, 1)$. As $\rho \rightarrow 1$,
the beta distribution essentially produces binary data with
$p = \mu$\footnote{Note that with $\rho < 1$, the beta distribution
  can still produce near boundary values, i.e. values arbitrarily close
  to zero and one. If at least one of $\alpha$ or $\beta$ is less than
  one, this can occur.}; smaller values of $\rho$ produce more
values within the interval $(0,1)$.
This sets the mean and variance of the beta distribution to
$\mu$ and $\sigma^2$ respectively.
Alternatively, we generate uniform noise bounded to keep $Y$ in the
unit interval.

For each simulated model, we also run an ``oracle'' model: we
enter the random vector $\bm{\cB}$ into this model as a fixed
predictor. This allows us to fit the model with
the \emph{glm}
function's \emph{quasibinomial} distribution family, from \textbf{R}.

\subsection{Type I Errors}

Our primary interest is verifying correct
type I coverage. That is, if the true effect
is zero, our 95\% confidence intervals should
cover zero at least 95\% of the time.

\begin{figure*}[ht!]
  \centering
  \includegraphics[width = 135mm]{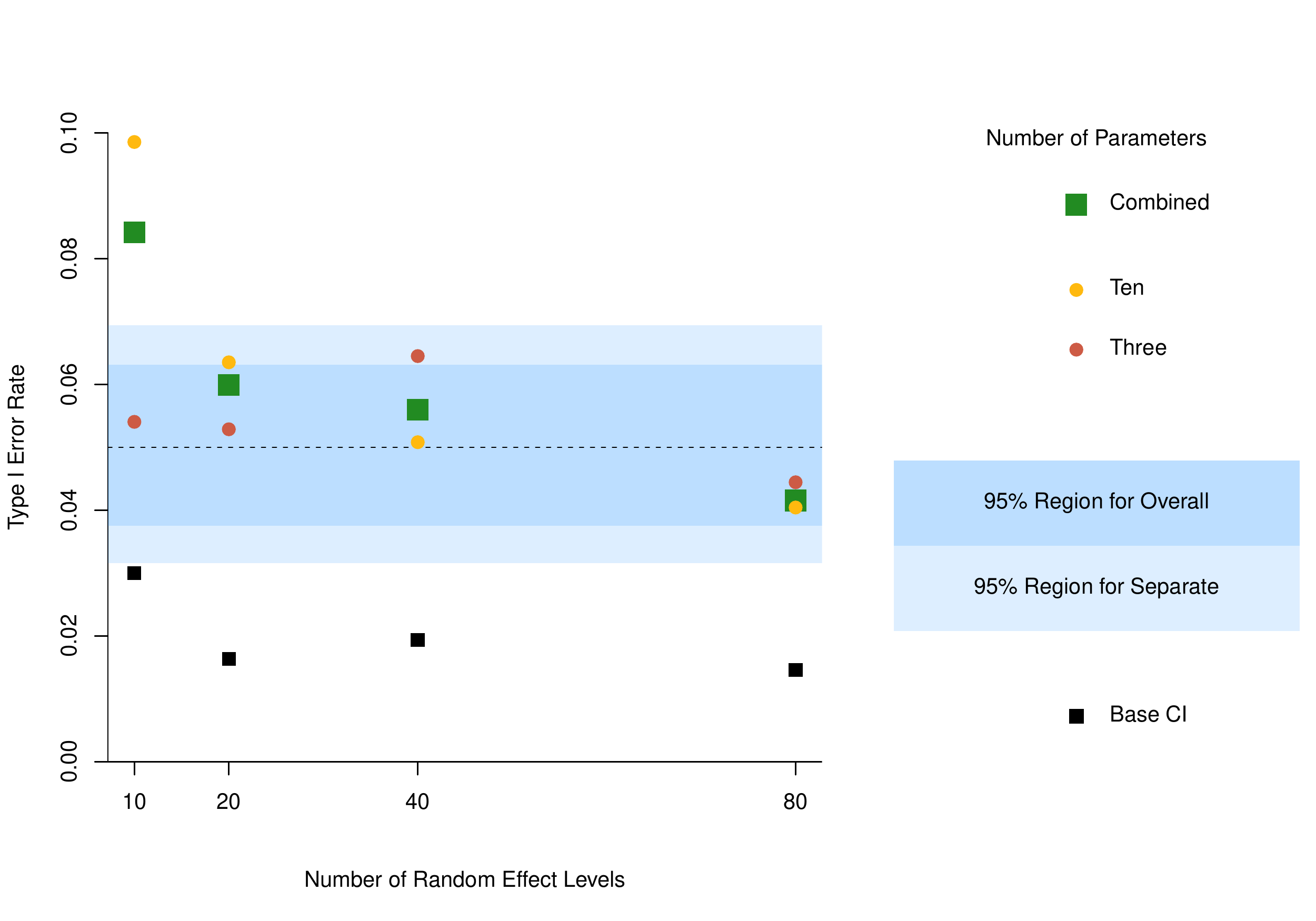}  
  \caption{Type I error by number of levels in the random effect. Green points are overall
    coverage for that level, yellow are for three-parameter models and red for the
  ten-parameter models. The black points below are from the base model. }
  \label{fig:one_rand}
\end{figure*}

We first plot the results with a single random effect.
Figure \ref{fig:one_rand} plots the type I error
against the number of levels in the random effect,
both overall in green, and separated by number
of paramters in the model in yellow and red, for ten and three
parameters respectively. The light blue regions are the
95\% prediction intervals under the null; the inner regions for the
overall result, and the outer for the separated results\footnote{Larger region
  due to lower counts.}. Note that there are more ten-parameter
effects than three. Finally, we add the coverage rates of the
base model confidence interval in black.

In general, these results are good. Only one result stands
out as poor: when we have ten levels and ten fixed effects,
not including the intercept. This is a difficult case for any
non-linear model. Once we reduce our model to just
three fixed effects beyond the intercept, even
with ten levels in the random effect we get correct coverage.

In constrast, the base model is very conservative, rejecting
only about $2\%$ of the time from a nominal $5\%$.

\begin{figure*}[ht!]
  \centering
  \includegraphics[width = 135mm]{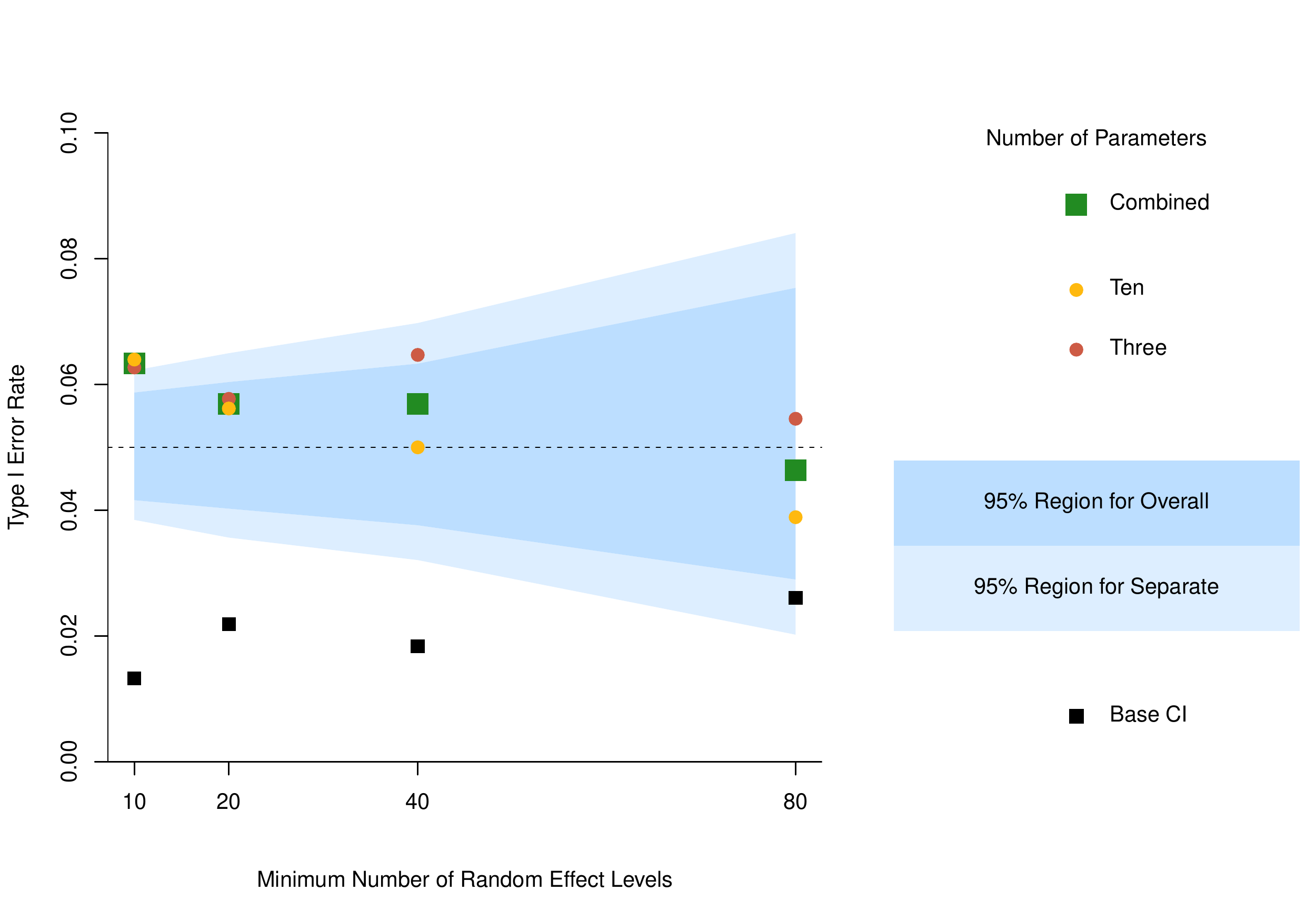}  
  \caption{Type I error by number of levels in the random effect. Green points are overall
    coverage for that level, yellow are for three-parameter models and red for the
  ten-parameter models. The black points below are from the base model. }
  \label{fig:two_rand}
\end{figure*}

Figure \ref{fig:two_rand} shows the results for two random effects. While
we bootstrap the effect with more levels, the effect with fewer levels has a stronger
effect on coverage. Thus we plot coverage against the smaller of the two level counts.
From there, the only major difference relative to the first plot is that the
number of elements to check is not the same at each level.

The results are good here too - for all level counts except ten,
all coverage intervals are within their 95\% null intervals. And again,
the base intervals are generally quite conservative.

\subsection{Relative Interval Length}

It's worth seeing how the interval widths of the bootstrapped intervals
compare to the base interval widths. Recall above we control the
standard deviation of $Y$ with a parameter $\rho$, and
as $\rho \rightarrow 1$, we get binary data; smaller values of $\rho$
give more values near the expected mean of $Y$. Thus we plot the relative width
of the intervals, i.e. bootstrap width divided by base width,
against the parameter $\rho$.

\begin{figure}[ht!]
  \centering
  \includegraphics[width = 85mm]{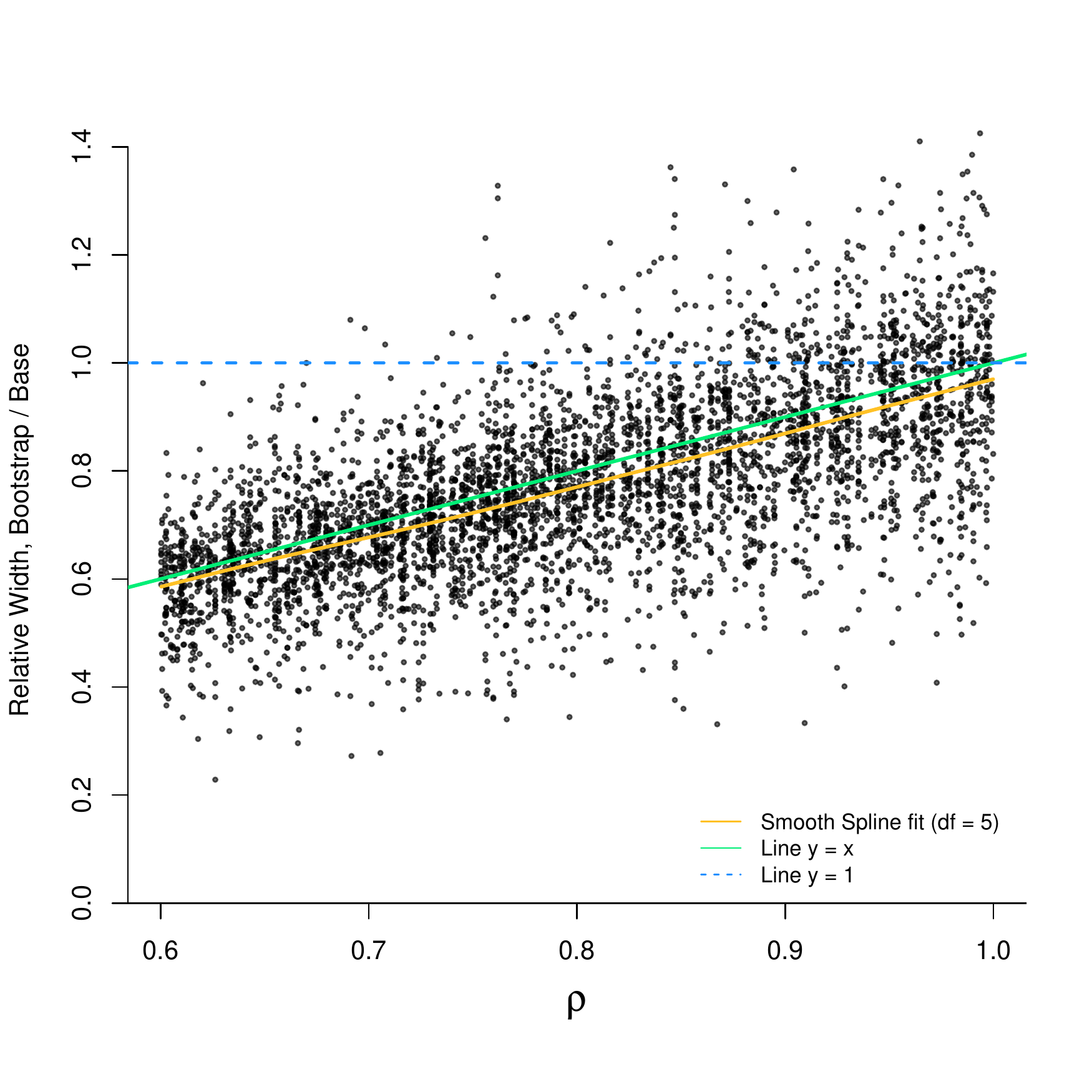}  
  \caption{The ratio of the bootstrap intervals' widths to the base
    interval widths, along with a smooth spline fit, and the
  theoretical line of equality.}
  \label{fig:interval_width_sd}
\end{figure}

Figure \ref{fig:interval_width_sd} plots the ratio of bootstrap width divided
by base width against the values of $\rho$ we use in our study. We fit a
smooth spline, in gold, and the line of equality $y = x$, in green, and we add the
line $y = 1$ in blue for comparison.

The smooth spline, with five degrees of freedom, is nearly linear and the
line is just below the line of equality. This suggest that on
average, the bootstrap is doing the right thing: it scales the
length of the intervals down by the same amount the standard deviation
of the $Y$ values are relative to the binary case. In fact, it's smaller
than that, as we can see with the fit line below the line of equality everywhere.

Note that this could give a shorthand method of getting confidence intervals from
regression methods: just scale down the width of the interval by $\rho$. Unfortunately,
we don't know $\rho$; seeing boundary values could mean the expected value of $Y$ was close
to zero and the standard deviation was small, or it could be a moderate expected $Y$ and a large
$\rho$. Further, there won't be a fixed $\rho$ for real data.

\section{Conclusion}
We have presented a method for analyzing bounded continuous outcome data in which outcomes near the bounds are frequent, but data within the range are also common. Under such conditions simply performing an ordinary linear regression is inappropriate because such analyses presuppose normally distributed error terms, resulting in
potentially biased inference.
Another common technique, that of collapsing outcomes over some grouping variable (such as subject and condition), is not wrong in and of itself, but it may be underpowered, as it excludes entering trial-level predictors into the model, and suffers other defects as indicated above. The present procedure avoids these problems.  

The primary costs to implementing the present method are that (a) the outcomes are presented in possibly unfamiliar units, and (b) bootstrapping to give precision to the confidence intervals demands a more substantial set of computations.  To manage (a), we recommend providing explicit examples in the description of the model.  For example, the regression output table should include not only the ${\beta}$ coefficients, but also their exponent, perhaps accompanied by an illustrative example given in the text (e.g., ``exp($\beta$) estimates the multiplicative change in the ratio of outcome \textit{a} to outcome \textit{b} given a one-unit increase in [that variable]." followed by a specific numerical example).  Readers familiar with logistic regression will already be familiar with interpretation of odds ratios, but others might not.  

Managing the computing requirements of the bootstrap analyses depends on the researcher's computing resources, but note that in a normal analysis pipeline the bootstrapping analysis will generally follow model exploration; that is, establishing definitive confidence intervals comes after most of the analysis work is done.  This minimizes the practical costs of implementing the procedure, with the benefit of less conservative inference.

Code for the R platform, together with illustrative examples, is available through
the Comprehensive R Archive Network (CRAN), at \\\texttt{https://cran.r-project.org/web/packages/glmmboot}.


\printbibliography

\input{po_appendix.tex}

\end{document}

%% file: packages.tex
\usepackage[linesnumbered,lined,boxed,commentsnumbered]{algorithm2e}    
\usepackage{amsmath}
\usepackage{amssymb}
\usepackage{bbm}
\usepackage{bm}
\usepackage{caption}
\usepackage{colortbl}
\usepackage{comment}
\usepackage{dsfont}
\usepackage[shortlabels]{enumitem}
\usepackage{float}
\usepackage{graphicx}
\usepackage{grffile} 
\usepackage[utf8]{inputenc}
\usepackage{lipsum}
\usepackage{listings} 
\usepackage[framemethod=TikZ]{mdframed}
\usepackage{multirow}
\usepackage{pdfpages}
\usepackage{pgfplots}
\usepackage{scalerel}
\usepackage{siunitx}
\usepackage{subcaption}
\usepackage{tabularx}
\usepackage{tikz}
\usetikzlibrary{tikzmark}
\usepackage[normalem]{ulem}
\usepackage{xcolor}
\usepackage{xfrac}
\usepackage{lmodern}
\usepackage{wasysym} 
\usepackage{perpage} 


%% file: math.tex
\newcommand{\x}{{\mathbf{x}}}

\newcommand{\E}{{\mathbf{E}}}
\renewcommand{\P}{{\mathbf{P}}}

\newcommand{\td}{\; \mathrm{d}}

\newcommand{\cB}{\mathcal{B}}

\newcommand{\cN}{\mathcal{N}}

\newcommand{\approptoinn}[2]{\mathrel{\vcenter{
  \offinterlineskip\halign{\hfil$##$\cr
    #1\propto\cr\noalign{\kern2pt}#1\sim\cr\noalign{\kern-2pt}}}}}

%% file: misc.tex
\usetikzlibrary{matrix}

\restylefloat{table}



%% file: po_appendix.tex
\clearpage

\appendix
\section{Logistic Log Likelihood}
\label{appen_quasill}

Note again that in the binary case, we have
$\E[Y] = \P(Y = 1) \cdot 1 + \P(Y = 0) \cdot 0 = \P(Y = 1)$,
This makes
solving for $\bm{\beta}$ straight-forward, since we have the full
distribution of the data, and therefore the likelihood.

We can also write this probability in terms of any
given $y$. Raising any number to the power 1
leaves it unchanged, and raising any nonzero number to the
power zero gives unity, thus we can write:

\begin{equation}
\begin{gathered}
\P(Y = y \mid \x) = \\
\P(Y = 1 \mid \x)^y \times \P(Y = 0 \mid \x)^{1 - y} = \\
\P(Y = 1 \mid \x)^y \times (1 - \P(Y = 1 \mid \x))^{1 - y}
\label{eq:proby}
\end{gathered}
\end{equation}

Checking the two cases, $y = 1$ and $y = 0$, shows the equality.

We want to write the likelihood using this.
Likelihood of the data is
$L(\bm{\beta} \mid \bm{y}, \x) = P(\bm{Y} = \bm{y} \mid \bm{\beta}, \x)$.
For one trial, this is:
\[
L(\bm{b} \mid y_i,  \bm{x_{i}}) = \P(Y = y_{i} \mid \bm{\beta} = \bm{b}, \bm{x_{i}})
\]

In words, the likelihood of a parameter vector given the data is the
probability of the data
given that parameter vector, with both conditional on the predictors $X$.

The parameters that maximize the log likelihood are the same parameters that
maximize the likelihood. The log-likelihood is easier to maximize, since
it turns products into sums, thus we work with the log likelihood.
Let $l(\bm{\beta} \mid \bm{y}) = \log L(\bm{\beta} \mid \bm{y})$ be the log likelihood.
To keep our
notation easier, let $L_{i}(\bm{b}) = L(\bm{b} \mid y_{i}, \bm{x_{i}})$ and
$l_{i}(\bm{b}) = \log L_{i}(\bm{b})$ and we'll drop writing $\bm{x_{i}}$
wherever it's implied.

Using \eqref{eq:proby}, we can write:
\[
L_{i}(\bm{b}) = \P(Y = 1 \mid \bm{b})^{y_{i}}
(1 - \P(Y = 1 \mid \bm{b}))^{1 - y_{i}}
\]

To reiterate, this just means:
\begin{align}
  \begin{split}
  L_{i}(\bm{b}) =
  \begin{cases}
      \P(Y = 1 \mid \bm{b}) & \text{if } y_{i} =
      1\\ \P(Y = 0 \mid \bm{b}) & \text{if }
      y_{i} = 0
  \end{cases}
  \label{eq:probline}
  \end{split}
\end{align}

Using \eqref{eq:Gexp}, we can then write:

\begin{align}
  \label{eq:lcase}
L_{i}(\bm{b}) = G(\x_i' \bm{b})^{y_{i}} (1 - G(\x_i' \bm{b}))^{1 - y_{i}}
\end{align}

Similarly for the log likelihood, we write:

\begin{align}
  l_{i}(\bm{b}) =& \log(\P(Y = y_{i} \mid \bm{b})) \\
  \begin{split}
    =& \log(\P(Y = 1 \mid \bm{b})^{y_{i}}
    (1 - \P(Y = 1 \mid \bm{b}))^{1 - y_{i}}) \\
    =&\; y_{i} \log[\P(Y = 1 \mid \bm{b})] + (1 - y_{i})
    \log\left[1 - \P(Y = 1 \mid \bm{b})\right]
  \end{split}
\end{align}

Going through both cases, we can again see that this is simply:
\begin{align}
    \begin{cases}
      \log[\P(Y = 1 \mid \bm{\beta} = \bm{b})] & \text{if }
      y_{i} = 1\\ \log[\P(Y = 0 \mid \bm{\beta} = \bm{b})] &
      \text{if } y_{i} = 0
    \end{cases}
    \label{eq:lprobline}
\end{align}

We again use \eqref{eq:Gexp} to write:
\begin{align}
  \begin{split}
  l_{i}(\bm{b})
    =&\; y_{i} \log[G(\x_i' \bm{b})] + (1 - y_{i})
    \log\left[1 - G(\x_i' \bm{b})\right]
    \label{eq:lprobinterp_a}
  \end{split}
\end{align}

However, the above log likelihood function can be evaluated with
$y_{i}
\in [0,1]$, not just for $\{0,1\}$. We lose the direct interpretation
of equations \eqref{eq:probline} and \eqref{eq:lprobline}, but
plugging any $y_{i} \in [0,1]$ in for equation
\eqref{eq:lprobinterp_a} shows that we now have a linear interpolation
between the $0$ and $1$ cases from \eqref{eq:lprobline}.
This is no longer a true likelihood, but that is not required to
find a consistent estimate of $\bm{\beta}$, while assuming
\eqref{eq:Gexp}.

From here, we maximise the full quasi-log likelihood
$L(\bm{b}) = \prod_{i} L_{i}(\bm{b})$, and thus
we find the parameter vector that best fits the
expected proportion $Y$ given our predictors.

\section{Further Logistic Interpretations: Endpoints}
\label{sec:append_interp_log}

When the ratios are very small, the ratio of ratios will be very close
to the ratio of proportions. For example, assume we have two proportions
$\sfrac{1}{50}$ and $\sfrac{3}{100}$.
The ratio of proportions will be
$\sfrac{1}{50} \div \sfrac{3}{100}  = \sfrac{1}{50} \times \sfrac{100}{3} = \sfrac{100}{150} = \sfrac{2}{3}$,
or $\approx 0.667$.


Converting the proportions to ratios gives 1:49 and 3:97; dividing these as fractions produces
$\sfrac{1}{49} \div \sfrac{3}{97}$, or $\sfrac{97}{49 \times 3}$
or $\approx 0.660$. Thus, arithmetically the difference between comparing proportions, and comparing ratios, is small when both ratios are small themselves.

For an example when this approximation doesn't work, assume
we have two proportions $\sfrac{3}{4}$ and $\sfrac{1}{5}$.
The ratio of proportions will be $\sfrac{15}{4} = 3.75$.
Converting the proportions to ratios gives $3$ and
$\sfrac{1}{4}$ respectively, and thus the ratio
of ratios is $12$, not that close to $3.75$.


This is an arithmetic consequence
of the relationship between ratios and proportions.  Thus, with $\E[Y
  \mid \x]$ close to zero, a unit change in $x_j$ can be
interpreted as:
\begin{align}
  \frac{\widehat{\text{prop}_i}^\text{new}}{\widehat{\text{prop}_i}^\text{old}}
  \approx \exp(\hat{\beta}_j)
\end{align}
A bad team with expected possession 
of $5\%$ would expect their proportion to increase by a factor of $\exp(0.4) = 1.5$
with a one mm increase in rainfall, to get to $7.5\%$. This is close to the true value of $7.3\%$. 

Similary, when $\E[Y \mid \x]$ is close to one, i.e. large ratios, we can
interpret 
\begin{align}
  \frac{1 - \widehat{\text{prop}_i}^\text{new}}{1 - \widehat{\text{prop}_i}^\text{old}}
  \approx \exp(-\hat{\beta}_j)
\end{align}

Now a great team with expected possession of $90\%$ would expect
their proportion of \emph{not} having possession to decrease by a factor of
$\exp(-0.4) = \sfrac{2}{3}$, down to $6.7\%$ from $10\%$. This is close to the
true value of $7\%$.

Note that for the ratios interpretation, there is no approximation required.

\section{Multiple Random Effects}
\label{multi_RE}

We first discuss two random effects, which can
be independent or crossed.

\textcite{mccullagh2000resampling} shows that no resampling method
exists that gives the correct variance estimate for crossed random
effects. However, \textcite{owen2012bootstrapping} outline methods
that are conservative and therefore valid, and are only mildly
conservative in most cases. We focus on the simpler method for two
\textcite{owen2007pigeonhole} pigeonhole
bootstrap. With multiple uncrossed effects, this method should be
close to unbiased.

We consider our data as a two-way random array, in that we have two
factors, i.e. two random factors over which we intend to generalize
the results of the model. We sample each factor independently: within
each factor, we resample with replacement, with equal probability, to
get a sample size that matches our data. For example, suppose we have
a dataset that comes from a set of human participants who respond to a
series of trials, where each trial makes use of one item from a small
set of items.  Iif we have $n$ trials, $K$ participants and $L$ items
(with $K, L < n$), a full resample of participants would be a sample
of size $K$, with each one randomly sampled from the population of $K$
participants. For any given boostrap resample, we are very likely to
have some datapoints not show up while others do so multiple times.

If participant $i$ shows up $A$ times in a given resample, and item
$j$ shows up $B$ times, we take all rows with participant $i$ and item
$j$ $AB$ times. There may be no rows with this combination, or there
could be many. If for example participant $i$ is sampled twice, item
$j$ is sampled three times, and there are five rows in the original
dataset with participant $i$ and item $j$, then this particular
resample will pick up each of those rows six times, for a total of
$30$ rows with that participant-item combination.

\begin{figure}[t!]
  \centering
  \includegraphics[scale=.43]{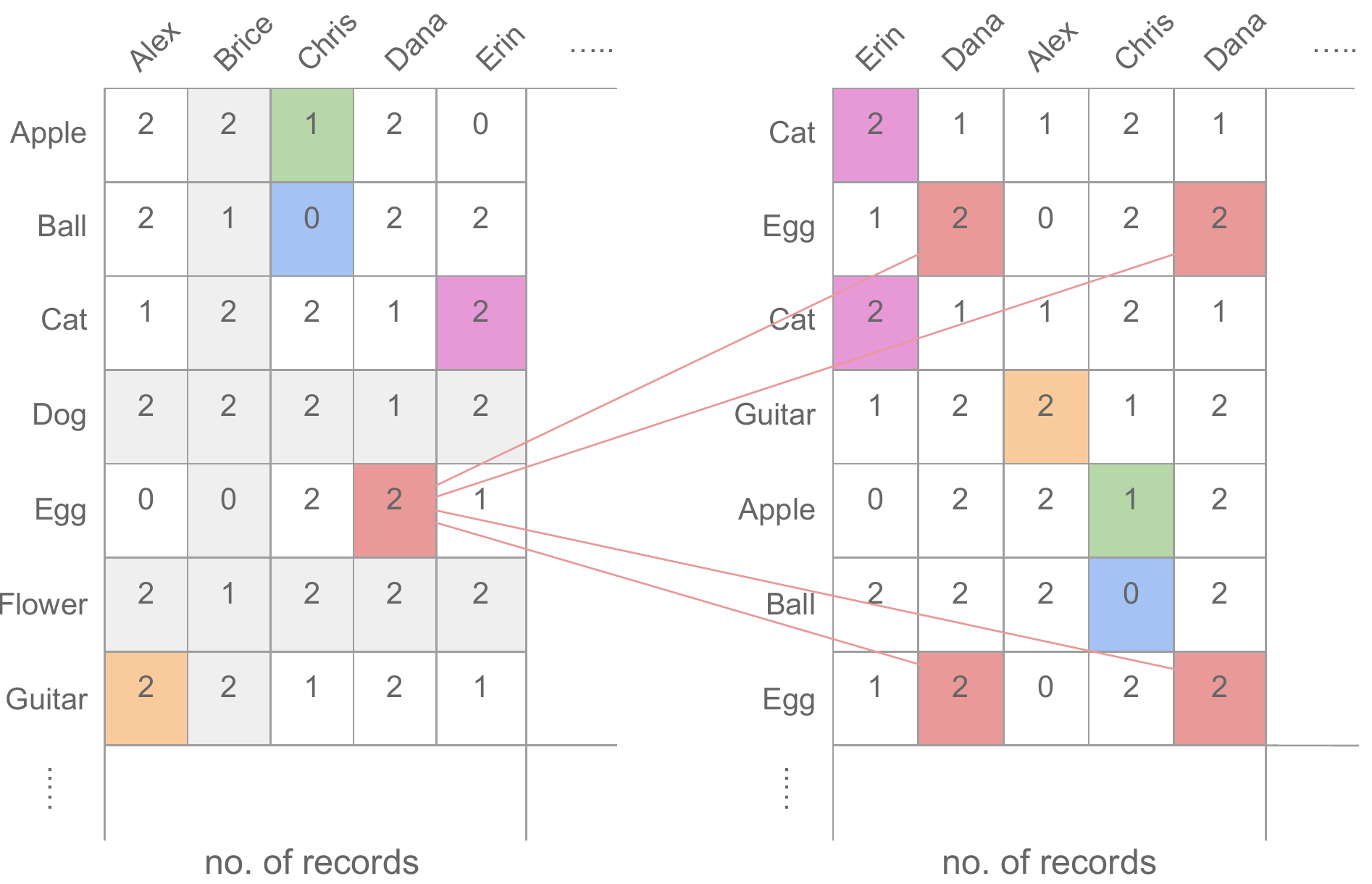}
  \caption{Our original data counts are on the left, and counts from a
    resampled data set are on the right. Matching colors are example cases of
    matched data counts.}
  \label{fig:resample}
  \vspace{\floatsep}
\end{figure}

Figure \ref{fig:resample} shows an example: our original data are on
the left, with a series of participants (named columns) responding to
a set of words (rows). We resample over items with replacement, and
over participants with replacement, to get our resampled data on the
right. Since data are resampled with replacement, we get some items
and participants showing up more than once, and some never. If a
pairing doesn't have any data in our original array, such as (Ball,
Chris), then there are no rows included for that pairing in our
resampled data either.

When we sample with replacement, each original value is selected at
least once with approximately a $0.63 = 1 - e^{-1}$ chance.  With a
two-dimensional array, each pair is selected at least once with an
approximately $0.4 = (1 - e^{-1})^2$ chance.


